# Graphene-Based Electrodes in a Vanadium Redox Flow Battery Produced by Rapid Low-Pressure Combined Gas Plasma Treatments

Sebastiano Bellani,*,▲ Leyla Najafi,▲ Mirko Prato, Reinier Oropesa-Nuñez, Beatriz Martín-García, Luca Gagliani, Elisa Mantero, Luigi Marasco, Gabriele Bianca, Marilena I. Zappia, Cansunur Demirci, Silvia Olivotto, Giacomo Mariucci, Vittorio Pellegrini, Massimo Schiavetti, and Francesco Bonaccorso*

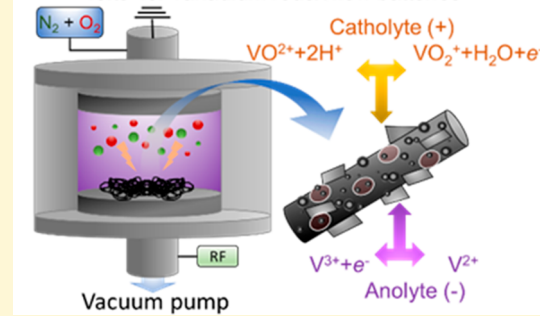

**ABSTRACT:** The development of high-power density vanadium redox flow batteries (VRFBs) with high energy efficiencies (EEs) is crucial for the widespread dissemination of this energy storage technology. In this work, we report the production of novel hierarchical carbonaceous nanomaterials for VRFB electrodes with high catalytic activity toward the vanadium redox reactions ($VO^{2+}/VO_2^+$ and $V^{2+}/V^{3+}$). The electrode materials are produced through a rapid (minute timescale) low-pressure combined gas plasma treatment of graphite felts (GFs) in an inductively coupled radio frequency reactor. By systematically studying the effects of either pure gases ($O_2$ and $N_2$) or their combination at different gas plasma pressures, the electrodes are optimized to reduce their kinetic polarization for the VRFB redox reactions. To further enhance the catalytic surface area of the electrodes, single-/few-layer graphene, produced by highly scalable wet-jet milling exfoliation of graphite, is incorporated into the GFs through an infiltration method in the presence of a polymeric binder. Depending on the thickness of the proton-exchange membrane (Nafion 115 or Nafion XL), our optimized VRFB configurations can efficiently operate within a wide range of charge/discharge current densities, exhibiting energy efficiencies up to 93.9%, 90.8%, 88.3%, 85.6%, 77.6%, and 69.5% at 25, 50, 75, 100, 200, and 300 mA cm$^{-2}$, respectively. Our technology is cost-competitive when compared to commercial ones (additional electrode costs < 100 € m$^{-2}$) and shows EEs rivalling the record-high values reported for efficient systems to date. Our work remarks on the importance to study modified plasma conditions or plasma methods alternative to those reported previously (e.g., atmospheric plasmas) to improve further the electrode performances of the current VRFB systems.

## 1. INTRODUCTION

Advanced large-scale energy storage systems (ESSs) are needed to meet the worldwide energy demand by exploiting renewable energy resources,[1−5] such as solar[6−8] and wind energies.[9−11] In fact, the intermittency and the instability of renewable power outputs have to be efficiently counterbalanced by the capability of ESSs to ensure a safe and reliable power supply continuously or on-demand.[12,13] In this context, redox-flow batteries (RFBs)[14−20] represent a promising stationary ESS technology because of their outstanding storage capability and output power[21−29] combined with prospective low costs,[30−38] easy scalability,[39,32,40] long lifetime,[41,42] low maintenance,[43,44] safety[44,45] and environmental friendliness.[44,45] Contrary to case-enclosed batteries, RFBs store the energy in the redox-active material-based electrolytes, filling external reservoirs.[14−18] The electrolytes flow from the reservoirs to the electrode surfaces, where the redox reactions occur rapidly compared to those in metal (e.g., Li, Na, K, etc.)-ion batteries.[46,46,47] As a result, the overall RFB capacities can be adapted to industrial-scale applications by enlarging the volume of the reservoirs independently by the power characteristics, which are defined by the size and number of cells in a module unit.[48−50,47] The energy density of a RFB is usually determined by three factors: (1) the concentration of the redox-active materials;[23,51−53] (2) the number of transferred electrons in the redox reactions;[54] and (3) the RFB voltage.[55,56] Among the RFBs, aqueous vanadium (V) redox flow batteries (VRFBs)[57−60] have been commercialized[61−64] thanks to their relevant energy and power performance coupled with the use of V-based species in both half-cells. The latter feature intrinsically diminishes the cross-contamination of active components,[65−67] eliminating the need of extensive separation techniques in order to recover the battery components at the end of the battery life, consequently lowering the costs of practical plants.[64,67] Vanadium also has the advantage of being recovered from industrial waste products, such as fly ash[68,69] or mine tailings,[70,71] cleaning up the environment. As a striking example, the "*World's largest battery*", a 200-MW, 800-MWh storage station based on VRFBs, manufactured by affiliated Rongke Power and UniEnergy Technologies (UET), is being built in the Dalian peninsula in northern China.[72]

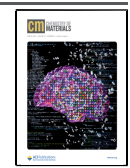

In order to encourage the market take-up of the VRFB technologies, researchers are struggling with improving the power density performance while retaining high energy efficiencies (EEs).[29,60,73–77] In fact, efficient high-power density operation can reduce the cell stack size,[74,75,78] consequently decreasing the capital cost of a VRFB plant.[74,75,73] Therefore, the development of feasible electrode materials with (1) high electrical conductivity,[79–81] which limits the ohmic polarization;[79–81] (2) high surface area with abundant catalytic sites for VRFB redox reactions (i.e., $VO^{2+}/VO_2^+$ and $V^{2+}/V^{3+}$ at the positive and negative electrodes, respectively);[78–81] and (3) hydrophilicity, which provides an optimal electrochemical accessibility of the redox-active materials to the electrode surface, is a research hotspot.[82] Nowadays, graphitic materials, in particular graphite felts (GFs), are regularly used as electrodes for commercial VRFBs[83–87] due to their low-cost manufacturing,[88] excellent electrical conductivity,[89,90] electrochemical stability,[89,85] and optimal hydraulic permeability.[89,89,91] However, their insufficient electrochemical activity toward the VRFB redox reactions[92,93] and low surface area (<1 m$^2$ g$^{-1}$)[89,90] severely limit the voltage efficiency (VE) and, thus, the overall EE of the VRFBs.[64,87,94] Furthermore, the hydrophobic nature of graphitic materials may hamper the electrolyte access to the electrode surface in the VRFBs.[84,92] Although the specific surface area can be simply increased by increasing the number of carbon fibers with reduced diameters, this strategy inevitably limits the hydraulic permeability of the electrode, increasing the energy required for adequate electrolyte pumping. To circumvent the trade-off between the specific surface area and the hydraulic permeability, several chemical/physical[76,84,95–100] and thermal treatments[29,92,101–103] have been reported to enhance the native electrochemical performance of the GFs. These approaches aim to introduce catalytic sites, such as chalcogen (O[92,95,99,100,104–109] and S[110]), N,[101,105,111–113] P,[114] and halogen[114] functional groups, and/or to increase the specific surface area by etching processes.[97,115–120] However, the most effective processes often require a prolonged processing time, toxic, corrosive, and expensive chemicals, and/or a high temperature.[64,83] Therefore, alternative methods are pursued to scale-up the manufacturing of highly efficient electrodes for commercial applications.[64,83] The incorporation of metals (Ir,[121,122] Au,[123] Pd,[123] Pt,[123,124] Cu,[125] and Bi[29,126–128]), metal oxides ($Nb_2O_5$,[129] $CeO_2$,[130,131] $ZrO_2$,[132,133] $PbO_2$,[134] $Mn_3O_4$,[135] $MoO_2$,[136] $Ta_2O_5$,[137] $Nd_2O_3$,[138] $NiO$,[139] and $WO_3$[140]) and inorganic pigments (e.g., Prussian blue)[141] into GFs as electrocatalysts has been also proposed. Nevertheless, the metals catalyze water splitting reactions,[121–124] while metal oxides show a limited electrical conductivity.[129,130,132,134–136,140] To bypass such drawbacks, carbon-based electrocatalytic nanomaterials,[142] including graphene derivatives,[143–155] carbon nanotubes,[156–166] carbon nanospheres/dots,[167–171] carbon black,[172] carbon nanosheets,[173] and carbon nanorods[174] have been used to decorate the GF surface. Although VRFBs with high rate capability (EE ≥ 80% at charge/discharge (CD), current densities ≥ 100 mA cm$^{-2}$)[29,143–145,147,164,170,171,174] have been successfully reported, either the cost or the long processing time of nanomaterial production and deposition hinder their practical implementation.[64,83] The detachment of nanomaterials can also negatively affect the long-term operation electrode performance, while contaminating the electrolyte.[64,83]

In this work, we report a rapid (minute time scale) production of texturized graphitic electrodes for VRFBs through a low-pressure combined gas plasma treatment of GFs in an inductively coupled radio frequency (RF) reactor. By systematically studying the effects of either pure gases, i.e., $O_2$ and $N_2$, or their combination, as well as the gas plasma pressure (set between 4 and 40 Pa), the electrodes were optimized to reduce their kinetic polarization toward VRFB reactions. To further enhance the surface area of the electrodes, single-/few-layer graphene, produced by the industrial wet-jet mill (WJM) exfoliation of graphite, were incorporated into GFs through a simple binder-aided infiltration method, dissolving polyvinylidene fluoride (PVDF) as the binder. After gas plasma treatment, the graphene-based electrodes showed a high rate capability. Our optimized VRFBs can efficiently operate in a wide range of CD current densities, from 25 mA cm$^{-2}$ (EE = 93.9%) to 300 mA cm$^{-2}$ (EE = 69.5%) with optimal cycling stability (over more than 200 cycles). Together with its low additional costs (<100 € m$^{-2}$) compared to commercial technologies, our electrode technology is market-competitive while showing EE values rivalling the current record-high values.

## 2. RESULTS AND DISCUSSION

### 2.1. Combined Multiple Gas Plasma Treatment of GFs.

In order to increase the electrochemically active surface area of the GFs without affecting the hydraulic permeability, combined multiple gas plasma treatments were investigated to attain a multiscale porosity while creating abundant catalytic sites through the incorporation of heteroatom functionalities. Briefly, the pristine GFs were treated by a combined $O_2$ and $N_2$ plasma using a $O_2$:$N_2$ (1:1 w/w) gas mixture in an inductively coupled RF (13.56 MHz) reactor. The gas plasma pressure was varied between 4 and 40 Pa to control the impact energy of the plasma species on the electrode surface, while fixing the plasma power and duration (see Experimental Section). As a comparison, some GFs were treated by either a single gas ($O_2$ or $N_2$) plasma step or by two sequential gas plasma steps ($O_2$ plasma followed by $N_2$ plasma or $N_2$ plasma followed by $O_2$ plasma). Hereafter, the electrodes treated by one gas plasma are named as X-P, in which X is the gas used for the plasma treatment (i.e., $O_2$, $N_2$, or $O_2$:$N_2$) and P is the applied gas plasma pressure (i.e., 4 Pa, 16 Pa, or 40 Pa). The electrodes treated by sequential plasmas are named X+Y-P, where X and Y are the gases used during the first and second gas plasmas, respectively. The proposed electrode treatments aim to provide a rapid alternative to the thermal processes commonly performed at high temperature (≥400 °C) for several hours (typically ≥6).[29,64,83] Notably, the proposed electrode modification is directly applicable within industrial VRFB supply chains. The rational of our strategy originated from the prior knowledge of standard gas plasma processes. In particular, the $O_2$ plasma creates reactive species, e.g., $O_3$ and O radicals, ionic species (e.g., $O^+$), and excited states thereof, which can react with carbonaceous surfaces, including the graphitic ones. For example, they can form O-based functionalities (e.g., hydroxyl (C–OH), carbonyl (C=O), and carboxyl (COOH) groups and aliphatic hydrocarbons),[175,84,176] which act as the catalytic sites for the VRFB redox reactions.[84,92,95,99,100,104−107,147] Moreover, morphological modifications of the surface of the carbonaceous materials can also occur during the O$_2$ plasma treatments, as a consequence of the C losses originated by either CO or CO$_2$ evolution.[84,100] Lastly, the O$_2$ plasma treatment is also effective for cleaning the carbonaceous electrode materials from organic contaminations.[177] Alternatively to the O$_2$ plasma, the N$_2$ plasma creates N atoms and radicals which effectively form N-based functionalities on carbonaceous surfaces.[178,179] For example, N$_2$ plasma treatment nitrates graphitic surfaces by generating C−N bonds,[180] thus introducing pyridinic-N, pyrrolic-N, quaternary-N, N-oxides of pyridinic-N, and aminic-N (more rarely graphitic N).[111,176,181,182] These functionalities have been demonstrated to be catalytically active for the VRFB redox reactions.[101,105,111−113,176] In addition, the five valence electrons of N atoms provide extra charges to the bond of the graphitic layers, enhancing their conductivity. Lastly, N$_2$ plasma can create structural defects, e.g., unsaturated C atoms, which react with either O$_2$ present in the electrode material or environmental O$_2$.[181] Although the aforementioned gas plasma treatments have been consecutively applied to modify GF for their use in VRFBs,[176] the combination of the O$_2$ and N$_2$ gases during the same plasma process has not been investigated yet. As we will show hereafter, the multiple plasma species in O$_2$:N$_2$ gas plasma give rise to synergistic effects in modifying the morphological, chemical, physical, and electrochemical properties of the GF surfaces, which can be engineered for the development of efficient VRFBs. The morphological modifications induced by the investigated plasma treatments were evaluated through scanning electron microscopy (SEM) measurements. Figure 1a shows the high magnification SEM image of a single fiber, which exhibits a smooth surface. The SEM image of a bundle of fibers in the pristine GF is shown in Figure S1. After the O$_2$ plasma at 40 Pa (O$_2$-40 Pa electrode), the surface of the fiber still shows a smooth surface (Figure 1b), which is similar to the one observed for the fibers in the pristine GFs. These results agree with previous literature,[84,100] in which no physical modifications were observed after O$_2$ plasma treatments (however, an excessive applied power, beyond the value here used, might etch the graphitic surfaces via a CO and/or CO$_2$ evolution reaction). Contrary to the O$_2$ plasma, the N$_2$ plasma at 40 Pa (N$_2$-40 Pa electrode) increases the coarseness of the fibers' surface compared to the one of the pristine GF fibers (Figure 1c). This morphology change is caused by the physical etching derived by N$_2$ plasma species impacting onto the GF surface, leading to structural defects (unsaturated C atoms).[181] These defects are highly reactive and are expected to react with O$_2$ plasma species, leading to both physical and chemical changes.[181] Indeed, the O$_2$:N$_2$ (1:1 w/w) plasma at the same pressure (O$_2$:N$_2$-40 Pa electrode) significantly enhances the coarsening of fibers' surfaces (Figure 1d). More in detail, the O$_2$ plasma species oxidize the fibers' surfaces, while the N$_2$ plasma species progressively etch the surface. The etching caused by the N$_2$ plasma species is promoted by the concomitant oxidation of the surface, while the oxidation is accelerated by the formation of structural defects.[181] Thus, the synergistic effects of the plasma species of two different gases foster "deeper" etching effects compared to the case of single gas plasmas. Beyond the gas plasma composition, the pressure of the plasma can significantly affect the morphology and chemistry of the final GF surface. More in detail, the mean free path between plasma species decreases with decreasing the plasma pressure. Consequently, the lower is the pressure, the longer is the mean acceleration time of the plasma species, which then impacts the surface of the treated sample with higher energy, boosting both the chemical modifications and the physical etching. Indeed, the O$_2$:N$_2$ (1:1 w/w) plasma at 16 Pa (O$_2$:N$_2$-16 Pa electrode) and the O$_2$:N$_2$ (1:1) plasma at 4 Pa (O$_2$:N$_2$-4 Pa electrode) lead to a multiscale texturization of the fibers' surfaces (Figure 1e,f), which shows (1) crater-like cavities with diameters ranging from a few hundreds of nanometers to above 1 $\mu$m (microtexturization) and (2) abundant and uniformly distributed micropores resulting in a nanoparticle-like appearance (nanotexturization).

The microtexturization is significantly more pronounced in O$_2$:N$_2$-4 Pa compared to O$_2$:N$_2$-16 Pa, as expected by the above discussion of the gas plasma processes. As shown by the subsequent electrochemical characterization, the multiscale texturization of the electrodes ensures an optimal hydraulic permeability by maintaining the macroscopic pathways for the electrolyte flow exhibited in the pristine GFs, while providing an elevated catalytic surface area for carrying out the redox reactions.

As above-discussed, the surface chemistry plays a crucial role in determining the catalytic activity of the VRFB electrodes.[183,184] Therefore, X-ray photoelectron spectroscopy (XPS) measurements were performed to evaluate the functional groups formed on the GF surface during the gas plasma treatments. The XPS wide scans of the various electrodes are reported in Figure S2, while the high-resolution spectra of the regions of C 1s, O 1s, and N 1s are shown in Figures S3−S5, respectively. As shown in Figure 2a, all the plasma treatments significantly increase both O and N functionalities. The O$_2$:N$_2$-4 Pa electrode shows the maximum O relative atomic percentage (at. %) of 15.9%, followed by O$_2$:N$_2$- 40 Pa and O$_2$:N$_2$-16 Pa (15.2 and 14.2%, respectively). Importantly, such O at. % values are significantly higher than those reported in the literature for thermally treated GFs optimized for VRFBs (typically lower than 8%).[29,76]

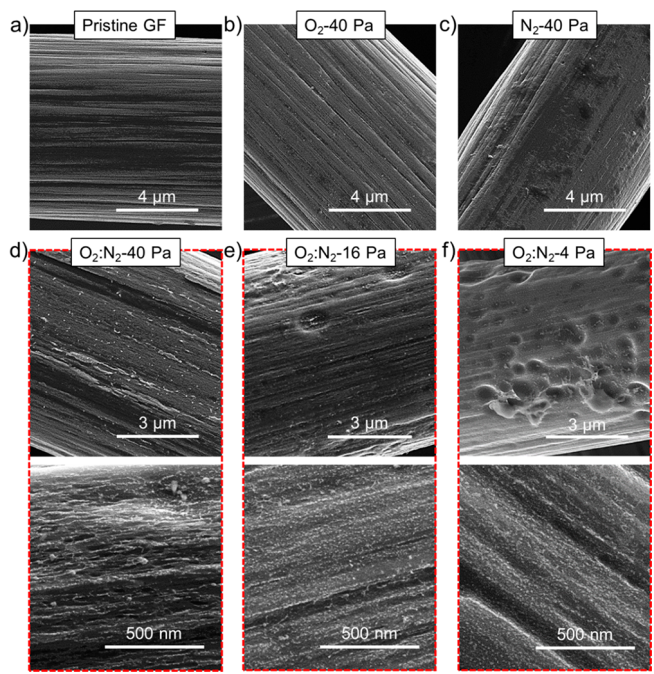

**Figure 1.** Morphological analysis of the pristine and plasma-treated GFs. SEM images of (a) pristine GF; (b) O$_2$-40 Pa; (c) N$_2$-40 Pa; (d) O$_2$:N$_2$ 40 Pa; (e) O$_2$:N$_2$ 16 Pa; and (f) O$_2$:N$_2$ 4 Pa. Panels (d)−(f) include two panels with different magnifications.

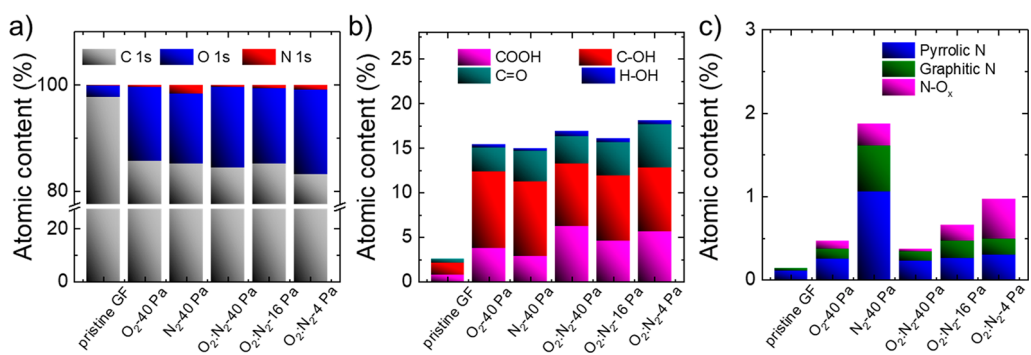

**Figure 2.** Chemical characterization of the pristine and plasma-treated GFs. (a) Elemental composition of the electrodes. (b) O and (c) N functionality distributions of the electrodes. The data have been estimated from the analysis of the XPS spectra (wide scan, C 1s, O 1s, and N 1s spectra).

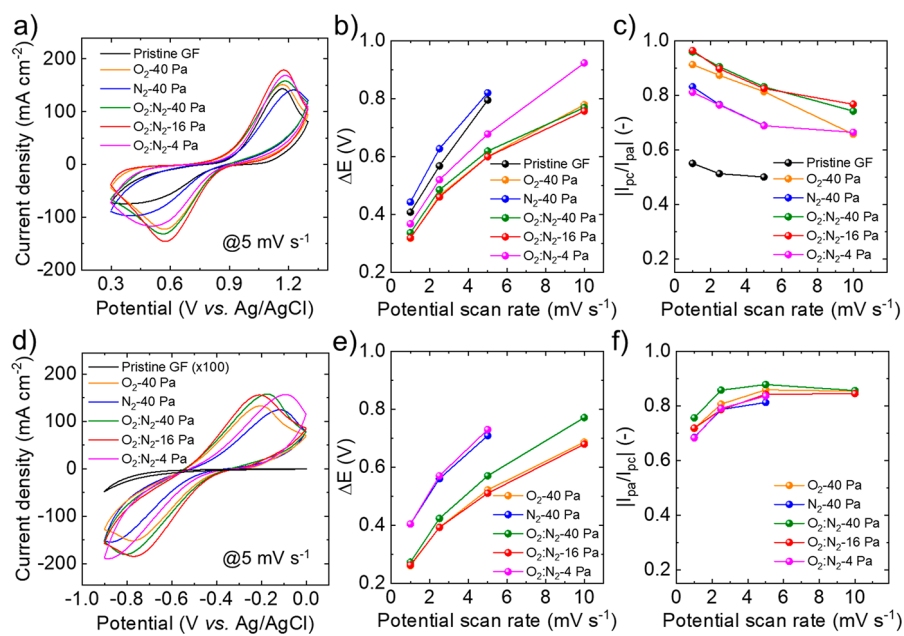

**Figure 3.** Electrochemical characterization of the pristine and plasma-treated GFs. (a) CV curves measured for the investigated electrodes in the 0.1 M VOSO$_4$ + 3 M H$_2$SO$_4$ solution for the VO$^{2+}$/VO$_2^+$ (positive) VRFB reaction at a scanning potential rate of 10 mV s$^{-1}$. (b) Analysis of $\Delta E$ and (c) $I_{pc}/I_{pa}$ (extrapolated by the CV curves shown in panel a). (d) CV curves measured for the investigated electrodes in the 0.1 M VOSO$_4$ + 3 M H$_2$SO$_4$ solution for the V$^{2+}$/V$^{3+}$ (negative) VRFB reaction at a scanning potential rate of 10 mV s$^{-1}$. (e) Analysis of $\Delta E$ and (f) $I_{pa}/I_{pc}$ (extrapolated by the CV curves shown in panel d). The current densities in (a) and (d) were calculated by normalizing the measured currents to the planar area of one face of the electrodes.

Thanks to the surface selectivity of the plasma treatments, a high surface oxidation of the GF can be obtained without altering the bulk properties, resulting in the optimal electrochemical performance (as we will show below). The maximum at. % of N is found for N$_2$-40 Pa (1.7%), followed by O$_2$:N$_2$-16 Pa and O$_2$:N$_2$-4 Pa (0.7% and 0.9%, respectively). As shown in the C 1s spectra, all the plasma treatments decrease the at. % of the C=C states and their satellite feature ($\pi$–$\pi^*$ peak), which means that the graphitic states (sp$^2$ hybridization) are converted to other states, including sp$^3$ hybridized states (which are mainly ascribed to the amorphous carbon[185] or carbon adatoms[185,186]), vacancy-like defects (which likely refer to plasma etching-induced structural modifications, e.g., pentagon, heptagon, and octagon carbon rings[185−187]), and heteroatomic functional groups.

The distributions of the O and N functionalities for the investigated electrodes were evaluated by the analyses of the O 1s (Figure S4) and N 1s spectra (Figure S5), respectively. As shown in Figure 2b, the plasma treatments increase the at. % of the carbonyl groups (C=O), hydroxyl groups (C−OH), and carboxyl groups (COOH), the latter two not shown by pristine GFs. All these O functionalities have been proposed to be catalytic sites for the VRFB redox reactions.[76,84,92,95,99,100,104−107,147,188,189] Furthermore, the plasma treatments introduce N groups, namely pyrrolic N, graphitic N, and N oxides (N−O$_x$), as illustrated in Figure 2c. Differently from N-doped electrodes obtained through high-temperature-assisted chemical functionalization,[190] no pyridinic groups were observed in our case. As previously demonstrated, N functionalities can act as catalytic sites for the VRFB redox reactions beyond the O functionalities,[101,105,111−113] with which they may lead to synergistic catalytic effects.[105,176,191]

Since the VRFBs operate in aqueous media, their electrodes must exhibit an optimal water wetting to guarantee an elevated electrolyte accessibility to the catalytic sites, as well as to increase the hydraulic permeability. Water contact angle measurements show that the surface of the pristine GF is hydrophobic (water contact angle = 122.9° ± 4.9°). Differently, the plasma-treated electrodes exhibit a zero-water contact angle (see Movie S1).

The conversion of the GF surface from hydrophobic to hydrophilic is directly attributed to the introduction of the polar groups of either O or N functionalities.[192,193]

**2.2. Electrode Characterization.** Cyclic voltammetry (CV) measurements in a three-electrode cell configuration were carried out to evaluate the catalytic activity of the pristine GF and the plasma-treated GFs toward the VRFB redox reactions. The catalytic properties of the electrodes can be evaluated from the analysis of the separation of the potentials of the current density peaks for the redox reactions ($\Delta E$),[194−196] as well as from the corresponding ratio of the anodic/cathodic or cathodic/anodic current density peaks ($I_{pa}/I_{pc}$ or $I_{pc}/I_{pa}$).[194−196] The measurements were performed at potential scan rates ranging from 1 to 10 mV s$^{-1}$ in 0.1 M $VOSO_4$ + 3 M $H_2SO_4$ solution. Even though this electrolyte is substantially different from the those typically used for the full VRFB systems (typically ≥1 M $VO^{2+}$ + 3 M $H_2SO_4$ and ≥1 M $V^{3+}$ + 3 M $H_2SO_4$ for the starting catholyte and anolyte, respectively), such analysis can qualitatively compare the behavior of the electron transfer kinetics for the VRFB reactions among different electrodes. Figure 3a shows the CV curves measured for the investigated electrodes between 0.3 and 1.3 V vs Ag/AgCl, in which the $VO^{2+}/VO_2^+$ redox reaction occurs,[197,198] at a potential scan rate of 5 mV s$^{-1}$. The $O_2$:$N_2$-16 Pa electrode shows the highest anodic current density among the investigated electrodes, followed by the $O_2$:$N_2$-4 Pa electrode. This trend can be associated with the superior surface area of the $O_2$:$N_2$-16 Pa and $O_2$:$N_2$-4 Pa electrodes, as shown by SEM analysis (Figure 1). Figure 3b,c shows $\Delta E$ and $|I_{pc}/I_{pa}|$, respectively, measured for the investigated electrodes as a function of the potential scan rate. At 10 mV s$^{-1}$, no cathodic peaks were observed for the pristine GF and the $N_2$-40 Pa electrode within the selected potential range. The $O_2$:$N_2$-16 Pa electrode shows the lowest $\Delta E$ values (~0.32 and ~0.76 V at 1 and 10 mV s$^{-1}$, respectively), indicating low overpotential for the $VO^{2+}/VO_2^+$ redox reaction. The pristine GF shows the lowest $|I_{pc}/I_{pa}|$ values (<0.6), which means a poor reversibility of the $VO^{2+}/VO_2^+$ redox reaction.[194−196] The highest $|I_{pc}/I_{pa}|$ values are measured for the $O_2$:$N_2$-16 Pa electrode, followed by the $O_2$:$N_2$-40 Pa electrode. For these cases, the $|I_{pc}/I_{pa}|$ values are higher than 0.9 for the potential scan rate equal or inferior to 2.5 mV s$^{-1}$. Although the $O_2$-40 Pa electrode shows $\Delta E$ and $I_{pc}/I_{pa}$ similar to those of the $O_2$:$N_2$-16 Pa electrode at a low potential scan rate (i.e., ≤5 mV s$^{-1}$), its performances (in particular $|I_{pc}/I_{pa}|$) are inferior to those of the most performing electrodes with increasing the potential scan rate to 10 mV s$^{-1}$. These effects can negatively affect high-power operation of the VRFBs based on $O_2$-40 Pa electrodes (see the result in Section 2.3). Figure 3d shows the CV curves of the investigated electrodes between −0.9 and −0.0 V vs Ag/AgCl, in which the $V^{2+}/V^{3+}$ redox reaction occurs,[197] at a potential scan rate of 5 mV s$^{-1}$.

The $O_2$:$N_2$-40 Pa, $O_2$:$N_2$-16 Pa, and $O_2$:$N_2$-4 Pa exhibit the highest current densities, likely due to their large surface areas. No clear anodic and cathodic peaks were observed for pristine GFs, indicating marginal catalytic activity for the $V^{2+}/V^{3+}$ redox reaction. As shown in Figure 3e, the $O_2$:$N_2$-16 Pa electrode shows the lowest $\Delta E$ values (0.26 and 0.68 V at 1 and 10 mV s$^{-1}$, respectively). Figure 3f shows that the $|I_{pa}/I_{pc}|$ values at the potential rate of 10 mV s$^{-1}$ are similar for the $O_2$:$N_2$-16 Pa and $O_2$-40 Pa electrodes, indicating comparable reversibility of the $V^{2+}/V^{3+}$ redox reaction. By decreasing the potential scan rate, the $|I_{pa}/I_{pc}|$ values decrease, likely due to the occurrence of the hydrogen evolution reaction (which is however eliminated in VRFBs working in proper voltage windows that guarantee high Coulombic efficiency, CE).[199] Importantly, the electrodes displaying a lower content of N functionalities (i.e., $O_2$-40 Pa and $O_2$:$N_2$-40 Pa, see Figure 2c) exhibit the highest redox reaction reversibility (i.e., $|I_{pa}/I_{pc}|$ values). This trend suggests that the N functionalities are the catalytic sites for the hydrogen evolution reaction, in agreement with previous literature.[200,201] Noteworthy, the hydrogen evolution reaction may be hindered in VRFBs operating at higher electrolyte concentration lowering the overpotential for the $V^{2+}/V^{3+}$ redox reaction.[199] Figure S6 shows $I_{pa}$ and $I_{pc}$, respectively, measured for the investigated electrodes as a function of the square root of the potential scan rate, for both the $VO^{2+}/VO_2^+$ and the $V^{2+}/V^{3+}$ redox reactions. The linear behavior of the curves indicates that the redox reactions are limited by the transport of the reactants toward the electrode surface, in agreement with the Randles−Sevick equation.[194−196,202] Moreover, the slopes of the $I_{pc}$ and $I_{pa}$ vs (potential scan rate)$^{1/2}$ plots of the gas plasma treated electrode are higher than the one of the GF (measured only for $VO^{2+}/VO_2^+$ redox reactions). This means that the chemical and morphology modification of the GF through the gas plasma treatment positively affect the reactant transfer rate toward the catalytic sites of the electrodes.[145,203]

**2.3. Evaluation of the Plasma-Treated Electrode-Based VRFB Performance.** The plasma-treated GFs were evaluated in VRFBs using a no-gap serpentine architecture,[204,205] Nafion 115 (thickness of 127 $\mu$m) as the proton exchange membrane, and 1 M $VO^{2+}$ + 3 M $H_2SO_4$ and 1 M $V^{3+}$ + 3 M $H_2SO_4$ as the starting catholyte (positive electrolyte) and anolyte (negative electrolyte), respectively. Hereafter, the VRFBs are named with the nomenclature used for their electrodes. First, polarization curve analysis was performed to evaluate the kinetic activation polarizations (kinetic losses) and the ohmic polarizations (iR losses) of the cells.[79,80] Figure 4a shows the *iR*-corrected polarization curves measured for the investigated VRFBs, specifically evidencing the kinetic losses resulting from the catalytic activity of the electrodes toward the VRFB redox reactions.[79,80] The raw polarization curves, which include the *iR* losses attributed to the resistance of the proton exchange membrane, the bipolar plates, and the current collectors, are reported in Figure S7. Clearly, the VRFBs based on plasma-treated electrodes significantly decrease the kinetic losses of the cell based on pristine GFs, in agreement with the CV data. The $O_2:N_2$-40 Pa VRFB shows the lowest kinetic losses, e.g., ~0.048, ~0.095, and ~0.167 V at 50, 100, and 200 mA cm$^{-2}$, respectively. In addition, we point out that the use of electrodes treated with two different gas plasmas (i.e., $O_2+N_2$-16 Pa and $N_2+O_2$-16 Pa) results in VRFBs with kinetic losses higher than those of the VRFBs using electrodes treated by low-pressure combined gas plasma (see Figure S8). Galvanostatic charge/discharge (CD) analysis was carried out to evaluate the efficiency metrics (i.e., the CE, the VE, and the EE) of the investigated VRFBs. Figure 4b shows the CD curves (second cycle) for the investigated VRFBs at the current density of 100 mA cm$^{-2}$. The upper voltage limit was fixed to 1.6 V in order to avoid parasitic reactions (i.e., water splitting reactions), in agreement with the best practices provided in literature.[197] In agreement with our CV analysis, the $O_2:N_2$-16 Pa VRFB shows the best electrochemical performance, reaching a discharge specific capacity of 11.2 Ah L$^{-1}$. These values correspond to an electrolyte utilization (EU) of 83.6%, being the theoretical capacity calculated on the total volume of the electrolyte, including both catholyte and anolyte, equal to 13.4 Ah L$^{-1}$. Figure 4c reports the CD curves measured for $O_2:N_2$-16 Pa VRFB at different current densities, ranging from 25 to 100 mA cm$^{-2}$. As expected, the capacity increases with decreasing the current density because of the reduced polarization losses. At 25 mA cm$^{-2}$, the capacity reaches values as high as ~12.5 Ah L$^{-1}$, corresponding to an EU of 93.6%. Figure 4d shows the rate capability of the $O_2:N_2$-16 Pa VRFB, showing the efficiency metrics over consecutive CD cycles at different current densities. At 100 mA cm$^{-2}$, the VRFB reaches a VE and an EE as high as 86.5% and 84.9%, which are among the highest values reported for VRFB using Nafion membranes with a thickness similar to our case (i.e., Nafion 115 or Nafion 117).[127,147,206] Similar or higher values have been recently reported using thinner Nafion membranes, e.g., Nafion 212 (thickness = 50.8 μm), Nafion 211 (thickness = 25.4 μm), or Nafion XL (thickness = 27.5 μm).[29,76,125] However, the latter systems are commonly tested at a current density superior to 100 or even 200 mA cm$^{-2}$, since they poorly perform during low-power density conditions due to the low CE (<95%) resulting by the cross-mixing of the vanadium species through thin Nafion membranes. Despite these issues, we anticipate that optimized VRFBs using a thin Nafion membrane (i.e., Nafion XL, thickness = 27.5 μm) will be shown later in the text for high power density applications. The main efficiency metrics of the VRFBs extrapolated by the galvanostatic CD measurements at various current densities are summarized in Table S1.

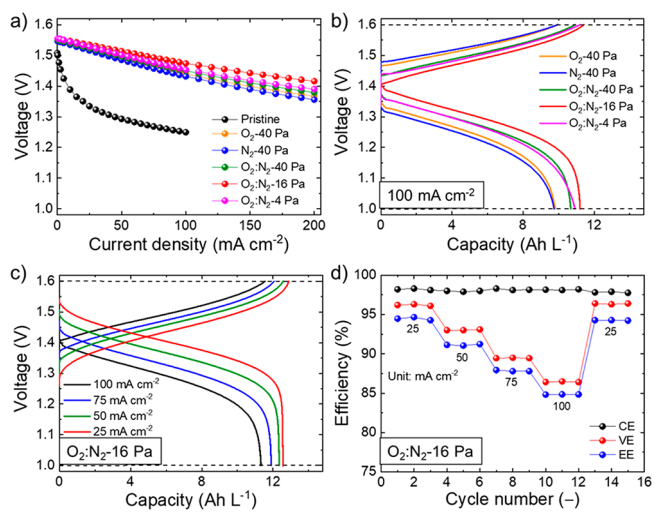

**Figure 4.** Electrochemical characterization of the VRFBs based on plasma-treated GFs using Nafion 115 as the proton-exchange membrane. (a) iR-corrected polarization curves measured for the VRFBs using pristine GFs or plasma-treated electrodes. (b) CD curves measured for the investigated VRFBs at a current density of 100 mA cm$^{-2}$. (c) CD curves measured for the $O_2$:$N_2$-16 Pa VRFB at various current densities (25, 50, 75, and 100 mA cm$^{-2}$). (d) Efficiency metrics (CE, VE, and EE) of the $O_2$:$N_2$-16 Pa VRFB using Nafion 115 extrapolated from the CD curve analysis as a function of the cycle number at various current densities.

## 3. APPLICABILITY OF THE GAS PLASMA TREATMENTS ON HIERARCHICAL GRAPHENE-COATED ELECTRODES

The applicability of the rapid combined plasma treatments was preliminarily evaluated for advanced hierarchical electrodes produced by decorating GF fibers with graphene flakes, aiming to increase the catalytic surface area of the GFs. Herein, hierarchical graphitic electrodes were produced by coating the GF with graphene. In order to maintain an industrial approach for the electrode fabrication, single-/few-layer graphene (SLG/FLG) flakes were produced through scalable wet-jet milling (WJM) exfoliation of graphite in *N*-methyl-2-pyrrolidone dispersion (see details in the Supporting Information).[207−209] Briefly, the WJM exfoliation process makes use of a high pressure (180−250 MPa) to transform a graphite dispersion in two jet streams, which then recombine in a small nozzle (diameter between 0.3 and 0.1 nm), where the generated shear forces cause the exfoliation of the graphite in single-/few-layer graphene flakes.[207,208,210] By applying three consecutive WJM passes on nozzles with diameters of 0.3, 0.15, and 0.1 nm, respectively, our WJM protocols lead to a highly concentrated dispersion (~10 g L$^{-1}$) of graphene flakes with an exfoliation yield of ~100% and a graphene production rate of ~2 g min$^{-1}$.[207,208,211] These values satisfy the requirements for high-throughput manufacturing chains of graphene-based commercial products.[212,213] The thorough characterization of the WJM-produced graphene flakes is reported in the Supporting Information (Figures S9 and S10). Importantly, as shown in previous works,[209] WJM-produced SLG/FLG flakes are pristine graphene flakes that do not exhibit basal plane defects, as also evidenced by Raman analysis (see Figure S10b). Consequently, they can guarantee superior electrical properties compared to other commercialized graphene derivatives, including graphene oxide and reduced graphene oxide. The hierarchical electrodes were produced by infiltrating the WJM-produced graphene dispersion mixed with polyvinylidene fluoride (PVDF) binder (weight percentage, wt % = 10%) into GFs (graphene mass loading of 20 mg cm$^{-2}$). The so-produced electrodes are herein named GF/graphene. Noteworthy, the graphene production rate of the WJM process is compatible with a production of 28 m$^2$ of electrodes per day. The additional material cost of our graphene-based electrodes is currently inferior to 100 € m$^{-2}$. Electrodes without the PVDF binder were also tested as a comparison to elucidate the adhesion effects between graphene and GF for the achievement of a durable electrodes performance. However, graphene flakes easily detached from the GF surface, leading to both poor data reproducibility and fast degradation of the VRFB performances during the electrochemical tests. As shown by the cross-section SEM images of a graphene/GF (Figures 5a,b), the GF fibers are wrapped by the graphene flakes, which also form clusters within the fiber-based felt network. The Brunauer−Emmett−Teller (BET) surface area of both the GF and the GF/graphene electrode was estimated by analyzing Kr physisorption measurements performed at 77 K.[214,215] The BET specific surface area of native GFs was ~0.4 m$^2$ g$^{-1}$, while it increased up to ~3.4 m$^2$ g$^{-1}$ (+750%) for GF/graphene.

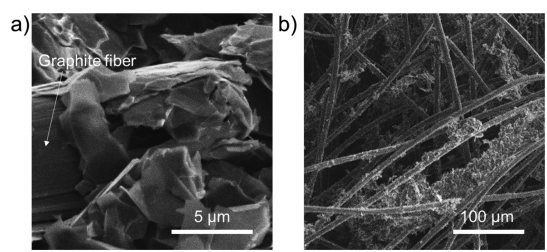

**Figure 5.** Morphological analysis of the GF/graphene electrodes. (a, b) Cross-sectional SEM images of the graphene-$O_2:N_2$-4 Pa electrode.

The low-pressure combined gas plasma treatments proposed for the pristine GFs were then applied to the so-produced GF/graphene, obtaining the electrodes herein named graphene-$O_2$:$N_2$-X Pa, in which X indicates the pressure of the gas plasma processes (i.e., 40, 16, or 4 Pa). Preliminary studies through polarization curve measurements on symmetric VRFBs also evaluated the electrochemical performance of GF/graphene electrodes treated with single gas plasma (i.e., $O_2$-40 Pa and $N_2$-40 Pa). Overall, our preliminary results indicated that the most performant VRFBs were those based on graphene-$O_2$:$N_2$-16 Pa and graphene-$O_2$:$N_2$-4 Pa, to which the discussion is directed hereafter. Noteworthy, the investigated gas plasma processes can also induce morphological and chemical modifications on the graphitic structure of WJM-produced graphene flakes, beyond the changes in the underlying GFs discussed in the previous section. However, the evaluation of the effects of the gas plasma parameters on the properties of the graphene flakes, as well as the optimization of the electrochemical performance of the corresponding graphene, is beyond the scope of this work and can be a subject matter of future studies. Figure S11 shows the polarization curves measured for the VRFBs using Nafion 115 as the proton-exchange membrane and based on low-pressure combined gas plasma-treated GF/graphene electrodes, namely, graphene-$O_2$:$N_2$-16 Pa and graphene-$O_2$:$N_2$-4 Pa, in comparison to the curve measured for the reference without graphene (i.e., $O_2$:$N_2$-16 Pa). Noteworthy, the incorporation of graphene flakes into the GF reduces the kinetic losses of the reference cell as a consequence of the increase of the number of electrode catalytic sites, likely introduced in the form of either O or N functionalities introduced by the plasma treatments.[216,217] The lowest polarization losses were measured for graphene-$O_2$:$N_2$-4 Pa. Figure S12 shows the comparison between the CD curves (second cycle) measured for the optimized VRFBs with and without graphene (i.e., $O_2$:$N_2$-16 Pa and graphene-$O_2$:$N_2$-4 Pa) at the current density of 100 mA cm$^{-2}$. The graphene-$O_2$:$N_2$-4 Pa VRFB exhibits the highest discharge capacity of 11.9 Ah L$^{-1}$, which corresponds to an EU of 88.8% (+6.2% compared to the graphene-free reference). Figure 6a reports the CD curves measured for the graphene-$O_2$:$N_2$-4 Pa VRFB using Nafion 115 as the proton-exchange membrane at various current densities, ranging from 25 to 100 mA cm$^{-2}$. At the lowest current density of 25 mA cm$^{-2}$, the cell reached an EE as high as ∼93.9%. At 100 mA cm$^{-2}$, the cells still show a high EE of 85.1%, which is the result of a high VE (86.9%). To fully exploit the low kinetic activation polarization of our graphene-based VRFBs, a thin Nafion XL was used as the proton-exchange membrane to reduce the ohmic polarization losses of the Nafion 115. Nafion XL consists of a microporous polytetrafluoroethylene (PTFE)-rich support layer (∼10 μm) impregnated on both sides with dense Nafion layer (∼10 μm).[218] Thanks to this three-layer structure, Nafion XL membranes demonstrated superior performance compared to their unreinforced analogue,[219] since it combined the advantages of microporous PTFE as a mechanical reinforcement,[218,219] and thin (∼27.5 μm) Nafion membrane as a fast-ion transporting channel.[218−220] Figure S13 compares the polarization curves measured for graphene-$O_2$:$N_2$-4 Pa VRFBs using Nafion 115 or Nafion XL, evidencing that the impact of the ohmic polarization losses is significantly weakened for the case of Nafion XL. Figure 6b shows that the graphene-$O_2$:$N_2$-4 Pa VRFB using Nafion XL can efficiently operate at a current density as high as 300 mA cm$^{-2}$, at which it shows a VE and an EE of 70.3% and 69.5%, respectively. Figure 6c,d reports the rate capability of the graphene-$O_2$:$N_2$-4 Pa VRFBs using Nafion 115 and Nafion XL, respectively.

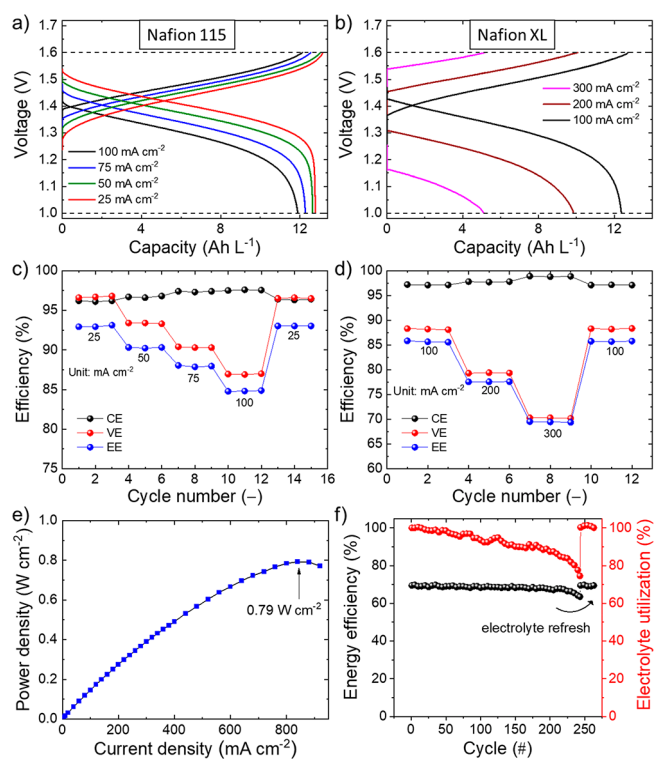

**Figure 6.** Electrochemical characterization of the VRFBs based on plasma-treated GF/graphene electrodes. (a) CD curves measured for the graphene-$O_2$:$N_2$-4 Pa VRFB using Nafion 115 at current densities of 25, 50, 75, and 100 mA cm$^{-2}$. (b) CD curves measured for the graphene-$O_2$:$N_2$-4 Pa VRFB using Nafion XL at current densities of 100, 200, and 300 mA cm$^{-2}$. (c, d) Efficiency metrics (CE, VE, and EE) of the graphene-$O_2$:$N_2$-4 Pa VRFBs using Nafion 115 and Nafion XL extrapolated from the CD curve analysis as a function of the cycle number at different current densities. (e) Power density as a function of the discharge current density measured for the graphene-$O_2$:$N_2$-4 Pa VRFB using Nafion XL. (f) Long-term stability tests of the CD performance of the graphene-$O_2$:$N_2$-4 Pa VRFB using Nafion XL at 300 mA cm$^{-2}$.

Table 1 summarizes the main efficiency metrics of the VRFBs extrapolated by the galvanostatic CD measurements at the investigated current densities. Clearly, Nafion XL enables the cell to efficiently operate at high power operation (i.e., current density > 100 mA cm$^{-2}$) thanks to its low resistance, which decreases the ohmic polarization losses. By benefiting from the low resistance of Nafion XL, the graphene-$O_2$:$N_2$-4 Pa VRFB can deliver a maximum power density as high as 0.79 W cm$^{-2}$ at the current density of 840 mA cm$^{-2}$ (Figure 6e). Meanwhile, Nafion 115 well performs at a current density ≤100 mA cm$^{-2}$, since it limits the electrolyte cross-mixing effects typically observed in thin Nafion,[221−223] guaranteeing superior CEs compared to Nafion XL.

**Table 1. Summary of the Efficiency Metrics of the Graphene-$O_2$:$N_2$-4 Pa VRFBs Using Nafion 115 or Nafion XL**[a]

| Nafion | current density (mA cm$^{-2}$) | CE (%) | VE (%) | EE (%) |
|---|---|---|---|---|
| 115 | 25 | 97.1 | 96.7 | 93.9 |
|  | 50 | 97.1 | 93.4 | 90.8 |
|  | 75 | 97.8 | 90.3 | 88.3 |
|  | 100 | 98.0 | 86.9 | 85.2 |
| XL | 100 | 97.1 | 88.2 | 85.64 |
|  | 200 | 97.7 | 79.4 | 77.57 |
|  | 300 | 98.8 | 70.3 | 69.45 |

[a]The values have been extrapolated from the galvanostatic CD measurements at various current densities and correspond to the second CD cycle.

Lastly, long-term cycling tests were performed to evaluate the durability of the proposed VRFBs. As shown in Figure 6f, graphene-$O_2$:$N_2$-4 Pa VRFB with Nafion XL optimally operates over more than 200 cycles without significant EE changes and a slight EU decrease (ca. −0.44%/cycle over the first 200 cycles). After such a number of cycles, the performance starts to decreases as a consequence of the anolyte losses and, thus, the change of the electrolyte composition,[220,222,224,225] caused by vanadium species/water permeability through Nafion XL. These effects are consistent with the water diffusion coefficient in Nafion,[226] which is between 10$^{-5}$ and 10$^{-6}$ cm$^2$ s$^{-1}$ (such values increase in the presence of cross-mixing of vanadium species).[227−229] The electrolyte refreshing, as typically needed to extend the stability tests over thousands of CD cycles,[171,29,120,168] restores the initial CD performances of the cells, indicating the electrochemical and mechanical stability of the plasma-treated graphene-based electrodes. At this stage, the development of strategies to limit anolyte losses is not the scope of this work, although it represents an utmost research topic in the field of VRFBs.[230−233] The exploitation of high-performant proton-exchange membranes[230,231,234,235] is expected to further extend the cycling performance of our VRFBs.

Figure 7 shows the comparison between the EE reached by our VRFBs and those reported in relevant literature. Clearly, our VRFBs based on low-pressure combined gas plasma-treated graphene-based electrodes exhibit EEs competing with those reported in the literature at both high and low current densities.[230,168,120,170,156,206,231,143,171,114,59,236,147,237] In particular, our VRFB EEs approach those achieved using advanced high-CE (∼100%) membranes, e.g., poly(ether sulfone)[230] and polybenzimidazole-based porous membranes,[231] as well as thin-film composite membranes based on an ultrathin polyamide selective layer on porous poly(ether sulfone)/sulfonated polyetheretherketone blend substrate.[235] Recently, high-pressure (atmospheric) plasma was also used to treat graphite felts, increasing the electrochemical performances of the reference VRFBs.[237] Consequently, remarkable EEs of 82.8%, 84.2%, and 77.5% were measured at the current densities of 120, 140, and 160 mA cm$^{-2}$, respectively.[237] Together with these results, our work remarks on the importance to study modified plasma conditions or methods to improve further the electrode performance obtained by means of the plasma treatments reported previously (including atmospheric plasmas).[99,237] In particular, our low-pressure plasma treatments were effective to improve the rate capability of our VRFBs, which reached EEs as high as 85.64%, 77.57%, and 69.45% at the current densities of 100, 200, and 300 mA cm$^{-2}$, respectively. Based on our previous characterizations, low-pressure plasma with combined $O_2$ and $N_2$ gases is effective to provide a multiscale texturization and a chemical O- and N-functionalization of our graphitic electrodes, which, thus, exhibit abundant catalytic groups for both the VRFB redox reactions. For the sake of completeness, we point out that two recent works reported by Zhao and co-workers have shown the EEs to be significantly superior to those achieved in any other system reported in the literature, reaching an EE higher than 70% at an almost incredible current density of 1 A cm$^{-2}$ (more than twice the maximum current investigated in the most relevant literature from other research groups).[29,76] These results are likely achieved thanks to an extra optimization of the system architecture beyond the design of efficient electrodes[29,76] and are therefore not included in Figure 7.

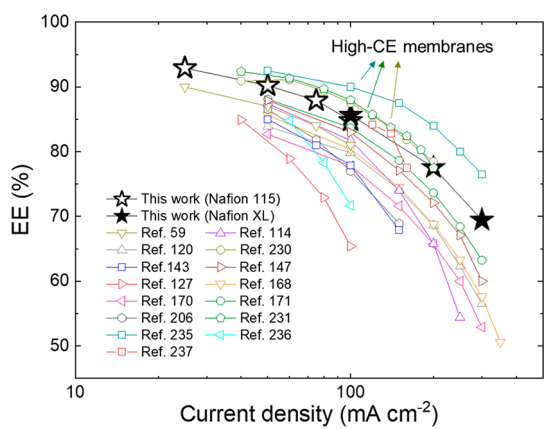

**Figure 7.** Comparison among the CD performances (EEs) of our graphene-based VRFB (graphene-$O_2$:$N_2$-4 Pa) and VRFBs reported in relevant literature.

## 4. CONCLUSION

We proposed a low-pressure combined gas plasma treatment in an inductively coupled radio frequency reactor to produce highly catalytic electrodes for vanadium redox flow batteries (VRFBs). The combination of multiple gases, namely, $O_2$ and $N_2$, in a single plasma process etches the surface of the GF fibers, while introducing both catalytic O and catalytic N functionalities. By investigating different gas plasma pressures, the electrodes were optimized to reduce their kinetic polarization toward VRFB redox reactions. The proposed low-pressure combined gas plasma treatments were further validated on hierarchical electrodes produced by decorating GF fibers with single-/few-layer graphene flakes produced through a scalable wet-jet milling exfoliation process. The optimized graphene-based VRFBs efficiently operate in a wide range of charge/discharge (CD) current densities, from 25 mA cm$^{-2}$ (energy efficiency = 93.9%) to 300 mA cm$^{-2}$ (energy efficiency = 69.5%) with optimal cycling stability over more than 200 cycles. Our VRFBs compete with the most efficient systems to date. Even more, our electrode technology has low additional costs (<100 € m$^{-2}$) compared to commercial ones, offering an industrial market-ready solution promoting the use of VRFBs for worldwide energy storage. Our work remarks on the importance to study modified plasma conditions or methods alternative to those reported previously (e.g., atmospheric plasmas) to improve further the electrode performances of the current VRFB systems. Prospectively, the use of advanced porous membranes with ultrahigh selectivity and stability could be used to boost the Coulombic efficiency of our VRFBs using Nafion membranes. In addition, the optimization of the architecture design could further decrease the kinetic losses, increasing the overall voltage efficiency during high-power density operation.

## 5. EXPERIMENTAL SECTION

**5.1. Graphene Production and Characterization.** The apparatus for the WJM exfoliation has been described in our recent patent[212] and studies.[207,238] Additional details regarding the fabrication and characterization methods are reported in Supporting Information, Methods.

**5.2. Formulation of the Graphene:PVDF Dispersion.** The WJM-produced graphene dispersion was purified by centrifuging at 1000 rpm for 30 min and collecting the supernatant. The so-produced dispersion was concentrated at 15 g L$^{-1}$ by evaporating N-methyl-2-pyrrolidone with a rotovapor at 60 °C. Then, PVDF (average molecular weight ∼534 000, Sigma-Aldrich) was added to the SLG/FLG dispersion in a material wt % of 10%.

**5.3. Electrode Production and VRFB Assembly.** The GF/graphene electrodes were fabricated by infiltrating 3 mL of the as-produced graphene:PVDF dispersion into GFs (4.6 mm GFD, Sigracell) with an area of 5 cm × 5 cm. Afterward, the electrodes were dried at 150 °C under vacuum for 1 h. Both pristine GF and GF/graphene were treated by combined multiple gas plasma, namely, O$_2$:N$_2$ plasma with a 1:1 (w/w) composition, in an inductively coupled radio frequency (13.56 MHz) reactor at a power of 100 W and a process pressure ranging from 4 to 40 Pa (background gas pressure of 0.2 Pa) for 10 min. Single gas plasma, namely, O$_2$ and N$_2$ plasma, and sequential single gas plasma (O$_2$ plasma followed by N$_2$ plasma, or N$_2$ plasma followed by O$_2$ plasma) were also investigated for comparison.

The VRFBs were assembled using a no-gap serpentine architecture (XLScribner RFB Single Cell Hardware). This hardware assembly consists of pairs of Poco Graphite flow-field layout-based graphite bipolar plates (Poco), Teflon flow frames, Viton rubber gaskets, and Au-plated Al end plates with electrolyte input/output ports (Swagelok fittings). Nafion 115 (thickness of 127 μm) (Dupont) was used as the proton exchange membrane. The as-produced electrodes were inserted into the space defined by the flow frames. A compression ratio of the electrodes of ∼30% was defined by the thickness of both the flow frames and rubber gaskets. Peristaltic pumps (Masterflex L/S Series) were used to flow the electrolyte into the cell hardware.

**5.4. Characterization of the Electrodes and VRFBs.** Water contact angle measurements were obtained by using DATAPHYSICS, OCA-15 setup, and Milli-Q water drops (2 μL) as the water reference. Scanning electron microscopy characterization was performed using a Helios Nanolab 600 and 450S Dual-Beam microscope (FEI Company) operating at 5 kV and 0.2 nA. The SEM images of the electrodes were collected without any metal coating or pretreatment. X-ray photoelectron spectroscopy (XPS) analysis was carried out using a Kratos Axis Ultra$^{DLD}$ spectrometer. The XPS spectra were acquired using a monochromatic Al Kα source operating at 20 mA and 15 kV. The analysis was carried out over an area of 300 μm × 700 μm. High-resolution spectra of C 1s, N 1s, and O 1s regions were collected at a pass energy of 10 eV and energy step of 0.1 eV. Energy calibration was performed setting the C−C peak in C 1s spectra at 284.8 eV. Data analysis was carried out with CasaXPS software (version 2.3.19).

Specific surface area analysis was carried out through Kr physisorption at 77 K[214,215] in Autosorb-iQ (Quantachrome). The specific surface areas were calculated using the multipoint BET model,[239] considering nine equally spaced points in a range of relative pressure ($P/P_0$, where $P_0$ is the vapor pressure of Kr at 77 K, corresponding to 2.63 Torr) between 0.10 and 0.30.

The electrochemical measurements were performed with a potentiostat/galvanostat (VMP3, Biologic). The CV measurements of the electrodes were carried in a three-electrode cell configuration using a KCl saturated Ag/ACl electrode as the reference electrode and a carbon rod as the counter electrodes. A 0.1 M VOSO$_4$ (>99.9%, Alfa Aeasar) + 3 M H$_2$SO$_4$ (ACS reagent, 95.0−98.0%, Sigma-Aldrich) solution were used as the electrolyte. The VRFBs were evaluated using 1 M VO$^{2+}$ + 3 M H$_2$SO$_4$ and 1 M V$^{3+}$ + 3 M H$_2$SO$_4$ as the starting catholyte and anolyte, respectively. The electrolytes were prepared from a 1 M VOSO$_4$ + 3 M H$_2$SO$_4$ solution through electrochemical method.[240] The electrolytes were pumped with a flow rate of 30 mL min$^{-1}$. Nitrogen was purged into the negative electrode reservoirs (containing V$^{2+}$ and V$^{3+}$) to avoid oxidation of V$^{2+}$ when the batteries were in a charged state. The polarization curve analysis was performed on charged VRFBs. The charged state of the VRFBs was reached by applying a constant current density of 100 mA cm$^{-2}$ and an upper voltage limit of 1.7 V. The VRFBs were then discharged for 30 s at each applied current density (ranging from 1 to 400 mA cm$^{-2}$, depending on the investigated cells). Cell voltage measurements were averaged over the 30 s of each current step to provide a point of the polarization curve. Before acquiring the polarization curves, the high-frequency resistance of the VRFB was measured by EIS at 30 kHz, in agreement with previously reported protocols.[79] The amplitude of the AC voltage perturbation was 10 mV. The iR losses were calculated by the product between the applied current ($i$) and the high-frequency resistance measured by EIS ($R$). The iR-corrected polarization curves were obtained by subtracting the iR losses from the raw polarization curves. The galvanostatic CD measurements of the VRFBs were carried out at different current densities (ranging from 25 to 400 mA cm$^{-2}$, depending on the investigated cells). The lower and upper cell voltage limits were set to 1 and 1.6 V, respectively.

## ■ ASSOCIATED CONTENT

**Ⓢ Supporting Information**

The Supporting Information is available free of charge at https://pubs.acs.org/doi/10.1021/acs.chemmater.1c00763.

> Water contact angle measurement of a plasma-treated electrode (AVI)
>
> Additional methods, SEM, TEM, AFM, Raman spectroscopy, XPS, and electrochemical characterizations (PDF)

## ■ AUTHOR INFORMATION

**Corresponding Authors**

**Sebastiano Bellani** − *BeDimensional S.p.a., 16163 Genova, Italy; Graphene Labs, Istituto Italiano di Tecnologia, 16163 Genova, Italy;* Email: s.bellani@bedimensional.it

**Francesco Bonaccorso** − *BeDimensional S.p.a., 16163 Genova, Italy; Graphene Labs, Istituto Italiano di Tecnologia, 16163 Genova, Italy;* orcid.org/0000-0001-7238-9420; Email: f.bonaccorso@bedimensional.it

**Authors**

**Leyla Najafi** − *BeDimensional S.p.a., 16163 Genova, Italy; Graphene Labs, Istituto Italiano di Tecnologia, 16163 Genova, Italy*

**Mirko Prato** − *Materials Characterization Facility, Istituto Italiano di Tecnologia, 16163 Genova, Italy;* orcid.org/0000-0002-2188-8059

**Reinier Oropesa-Nuñez** − *BeDimensional S.p.a., 16163 Genova, Italy; Department of Materials Science and Engineering, Uppsala University, 751 03 Uppsala, Sweden*

**Beatriz Martín-García** − *Graphene Labs, Istituto Italiano di Tecnologia, 16163 Genova, Italy; CIC nanoGUNE, 20018 Donostia-San Sebastian, Basque, Spain;* orcid.org/0000-0001-7065-856X

**Luca Gagliani** − *Graphene Labs, Istituto Italiano di Tecnologia, 16163 Genova, Italy*

**Elisa Mantero** − *BeDimensional S.p.a., 16163 Genova, Italy; Graphene Labs, Istituto Italiano di Tecnologia, 16163 Genova, Italy*

**Luigi Marasco** − *Graphene Labs, Istituto Italiano di Tecnologia, 16163 Genova, Italy*

**Gabriele Bianca** − *Graphene Labs, Istituto Italiano di Tecnologia, 16163 Genova, Italy; Dipartimento di Chimica e Chimica Industriale, Università degli Studi di Genova, 16146 Genoa, Italy*

**Marilena I. Zappia** − *BeDimensional S.p.a., 16163 Genova, Italy; Department of Physics, University of Calabria, 87036 Rende, Cosenza, Italy*

**Cansunur Demirci** − *Dipartimento di Chimica e Chimica Industriale, Università degli Studi di Genova, 16146 Genoa,*

*Italy; NanoChemistry, Istituto Italiano di Tecnologia, 16163 Genova, Italy*

**Silvia Olivotto** — *Wind Technology Innovation, Enel Global Power Generation,* https://www.enel.com/

**Giacomo Mariucci** — *Storage and New Business Design, Engineering & Construction, Enel Green Power S.p.A.,* https://www.enel.com/

**Vittorio Pellegrini** — *BeDimensional S.p.a., 16163 Genova, Italy; Graphene Labs, Istituto Italiano di Tecnologia, 16163 Genova, Italy*

**Massimo Schiavetti** — *Thermal & Industry 4.0 Innovation, Enel Global Power Generation,* https://www.enel.com/

Complete contact information is available at:
https://pubs.acs.org/10.1021/acs.chemmater.1c00763

### Author Contributions
▲(S.B. and L.N.) These authors contributed equally.

### Author Contributions
The manuscript was written through contributions of all authors.

### Funding
This project has received funding from Partnership Agreement Istituto Italiano di Tecnologia and ENEL Green Power S.p.A. (prot. IIT nr. 0018079/16). This project has received funding from the European Union's Horizon 2020 research and innovation program under grant Agreement No. 785219 and No. 881603-GrapheneCore2 and GrapheneCore3, European Union's MSCA-ITN ULTIMATE project under Grant Agreement No. 813036, European Union's SENSIBAT project under Grant Agreement No. 957273, and from the Italian Ministry of Foreign Affairs and International Cooperation (MAECI) through Cooperation Project "GINGSENG" (Grant PGR05249) between Italy and China.

### Notes
The authors declare no competing financial interest.

## ACKNOWLEDGMENTS
We thank the Electron Microscopy facility—Istituto Italiano di Tecnologia for the support in TEM data acquisition; Dr. Andrea Toma for the access to Clean Room facilities, Istituto Italiano di Tecnologia for carrying out the SEM analysis; and Dr. Riccardo Carzino (Smart Materials group at Istituto Italiano di Tecnologia) for the contact angle measurements.

## ABBREVIATIONS
CD, charge/discharge; CE, Coulombic efficiency; CV, cyclic voltammetry; EE, energy efficiency; ESSs, energy storage systems; GFs, graphite felts; PTFE, polytetrafluoroethylene; PVDF, polyvinylidene fluoride; RFBs, redox flow batteries; SEM, scanning electron microscopy; TEM, transmission electron microscopy; VE, voltage efficiency; VRFBs, vanadium redox flow batteries; WJM, wet-jet milling; XPS, X-ray photoelectron spectroscopy